\documentclass[english,aps,pra,showpacs,twocolumn,titlepage,longbibliography]{revtex4-1}   
\usepackage{geometry}		
\usepackage[above,below]{placeins}	
\usepackage{graphicx,color,braket,footnote,subfigure}
\usepackage{amsmath,amsthm,bm}
\usepackage[T1]{fontenc}
\usepackage[utf8]{inputenc}
\usepackage{amssymb}
\newcommand{\bla}{\color{black}}

\usepackage{hyperref}
\hypersetup{colorlinks,linkcolor={red},citecolor={blue},urlcolor={black}}
\usepackage{amsxtra}

\begin{document}
	\title{\bf Quantum causal correlations and non-Markovianity of quantum evolution} 
	 \author{Shrikant Utagi}  
	\email{shrik@ppisr.res.in}
	\affiliation{Poornaprajna    Institute   of   Scientific
		Research, Bengaluru -  562164, India} 
	\affiliation{Graduate  Studies, Manipal Academy of Higher Education, Manipal - 576104, India.}
	
\begin{abstract}
	A non-Markovianity measure for quantum channels is introduced based on causality measure - a monotone of causal (temporal) correlations - arising out of the pseudo-density matrix (PDM) formalism which treats quantum correlations in space and time on an equal footing. Using the well known damped Jaynes-Cummings model of a two-level system interacting with a bosonic reservoir at zero temperature as an example, it is shown that breakdown of monotonicity of the causality measure is associated with the revival of temporal (causal) correlations hence with the negativity of the decay rate. Also, a note on the comparison with other well known non-Markovianity measures is given.\\\\
	
	\textbf{\it Keywords:} Pseudo-density matrix, quantum temporal (causal) correlations, non-Markovianity
	\end{abstract}

\maketitle

	\section{Introduction}
Quantum non-Markovian dynamics has received significant attention over the last few years \cite{breuer2002theory,breuer2016colloquium}. Defining, characterizing and quantifying memory effects in open quantum system dynamics has been a rich topic of interest \cite{RHP14,shrikant2018non-Markovian}. 
A number of non-Markovianity measures, based on distance \cite{breuer2009measure,liu2013nonunital}, fidelity \cite{rajagopal2010kraus}, correlations \cite{RHP10,hou2011alternative,santis2019correlation}, channel capacity \cite{bylicka2014non}, canonical decay rate \cite{hall2014canonical} and that based on deviation from semigroup structure \cite{utagi2020temporal} have been proposed to name a few. However, equivalence of the measures is still under intense investigation \cite{li2018concepts}.

Lately, a number of approaches have been formulated to address quantum correlations in space and time in a unified picture \cite{zhang2020quantum,costa2018unifying}. A framework now known as pseudo-density matrix (PDM) formalism was introduced in \cite{fitzsimons2015quantum}, in which spatial and temporal correlations are taken into account in a unified way. Through a suitable mapping, the PDM formalism has been shown to be equivalent to a number approaches, e.g., process matrix formalism which deals with quantum theory without a definite global causal order although obeying quantum mechanics locally, for studying temporal correlations in non-relativistic quantum mechanics \cite{zhang2020quantum}. 

Recently, non-Markovian dynamics has been studied using temporal steering and quantified using temporal steerable weight \cite{chen2016quantifying}. Interestingly, a quantum channel with memory, arising due to correlation between successive use of channel does not violate causality: outputs of channel at $t$ do not depend on the input of the channel at $t'>t$ \cite{kretschmann2005quantum}. However, in the PDM formalism, application of channels preserves causality, but the causal order may be indefinite \cite{fitzsimons2015quantum}. See, e.g., \cite{zhang2020quantum}, where a tripartite PDM is shown to be equivalent to quantum switch - the only known higher order linear operation that gives rise to indefinite causal order \cite{ebler2018enhanced,chiribella2018indefinite}. Interestingly, violation of temporal Bell inequalities, i.e., temporal Tsirelsen bound, as an indication of completely positive (CP-) indivisibility was noted in \cite{le2017divisible}. In \cite{ku2018hierarchy} it was shown that temporal correlations due to temporal non-separability, temporal steering, and temporal non-locality form a hierarchy. 

In \cite{budini2018quantum}, a novel approach to non-Markovianity was given based on breakdown of conditional past-future correlations, in which the author proposes an indicator of non-Markovianity that detects memory effects that are due even to CP-divisible dynamics. In \cite{capela2020monogamy}, the notion of operational divisibility (where a non-Markovian process can be operationally divisible) has been associated with the satisfaction of data processing inequality and monogamy of temporal correlations. 

This work is focused on quantifying non-Markovianity using a recently introduced causality measure arising out of PDM formalism. Making use of \textit{causality monotone}, which is a measure of causal or temporal correlations, introduced in \cite{fitzsimons2015quantum,pisarczyk2019causal}, we will show that breakdown of its monotonicity is associated with non-Markovianity of a quantum channel. As we will show later, the measure introduced in this work is with regards to the processes that are CP-indivisible in the sense of \cite{RHP10}, which will be explained shortly, and positive (P-)indivisible in the sense of \cite{chruscinski2014degree}. Studying open system dynamics by considering spatial and temporal correlations on a equal footing may open up an arena to study memory effects with indefinite causal order and help understand a possible intimate connection between non-Markovianity and causal order \cite{milz2018entanglement}. We shall be using the terms temporal correlations and causal correlations (or relationships) interchangeably, since both mean the same in PDM formalism.  


\section{The pseudo-density matrix \label{sec:pdm}}
A density matrix can be given a Pauli operator representation $\rho = \frac{1}{2}\sum_{j=0}^{3}\alpha_j \sigma_j,$ with $\alpha_j = \langle \sigma_j \rangle = \text{Tr}(\rho \sigma_j)$, where $\sigma_0=I,\sigma_1=X,\sigma_2=Y,\sigma_3=Z$ are the Pauli operators. Similarly,  a two qubit joint state of two spatially separated systems $A$ and $B$ can also be given a Pauli representation as:
\begin{align}
\rho^{\rm AB} = \frac{1}{4}\sum_{i,j=0}^{3} \alpha_{ij} (\sigma_i \otimes \sigma_j),
\label{eq:paulibell}
\end{align} 
where $\alpha_{ij} = \langle \sigma_i \otimes \sigma_j \rangle = \text{Tr}(\rho^{\rm AB}\sigma_i \otimes \sigma_j)$ are the joint expectation values of two Pauli operators pertaining to symbols $i$ and $j$.  Here, the system density operators $\rho^A$ and $\rho^B$ act on different Hilbert spaces $\mathcal{H}^A$ and $\mathcal{H}^B$ respectively, each of dimension 2. When the tensor production structure in Eq. (\ref{eq:paulibell}) fails, one talks of a non-separable state; for example, entangled (Bell) states.

Now, we come to the framework to define \textit{correlations in time} by associating a tensor product structure to the qubit Hilbert spaces $\mathcal{H}_{t_1}$ and $\mathcal{H}_{t_2}$ to the density matrix $\rho_A$ at time $t_1$ and $\rho_B$ at time $t_2$ respectively. When such a tensor product structure fails, one talks about the temporal correlations \cite{ku2018hierarchy}. The time evolution is generally obtained by using a qubit completely positive trace preserving map $\mathcal{E}_{B \leftarrow A}$ such that $\rho_B = \mathcal{E}_{B \leftarrow A}[\rho_A]$. The state $\rho_A$ may be thought of as an input to the quantum channel $\mathcal{E}$. 
\bla
Now, this can be extended to $k$ sequence of Pauli measurements representing $k$-qubit state as below:
\begin{align}
\mathcal{P} = \frac{1}{2^k} \sum_{i_1 = 0}^{3} \cdots \sum_{i_k = 0}^{3} \langle \{\sigma_{i_j}\}^k_{j=1} \rangle \bigotimes^k_{j=1} \sigma_{i_j}
\label{eq:pdmgeneral}
\end{align}where $\langle\{\sigma_{i_j}\}^k_{j=1}\rangle = \text{Tr} \left[\left(\bigotimes\limits_{j=1}^{k}\sigma_{i_j}\right)\mathcal{P}\right]$ is the joint expectation value of $k$ Pauli observables where each measurement event is denoted by $j$ for each $i$. In other words it is nothing but the correlation function of a $k$ sequence of Pauli measurements $\sigma_{i_j}$ on the system \cite{zhao2018geometry}. Above state (\ref{eq:pdmgeneral}) shares all the features of a proper density matrix except that it may not be positive semi-definite.  However, the PDM is Hermitian and has unit trace. Since PDM is defined through the weighted sum over Pauli operators, the Hermiticity of Pauli operators giving rise to all real valued weights guarantees the Hermiticity of PDM. Unit trace of the PDM follows from the fact that Pauli operators are traceless and only trace of the identity operators contribute: $\text{Tr}[\mathcal{P}] = \text{Tr}[\frac{\langle I \cdots I \rangle}{2^k} I]= 1$. Even though, PDM is not a state globally (meaning correlated state with at least one negative eigenvalue), under the partial trace it gives a physically valid marginal state. For a simple case of two sequential measurements i.e., for $k=2$, the equation (\ref{eq:pdmgeneral}) reduces to \begin{align}
\mathcal{P}_{\rm AB} = \frac{1}{4}\sum_{i,j=0}^{3} \langle \sigma_i \otimes \sigma_j \rangle (\sigma_i \otimes \sigma_j).
\label{eq:pdm2point}
\end{align} 
It is important to note that PDM represents arbitrary mixtures of spatial and temporal correlations. Whenever PDM is not positive semi-definite, i.e., when it has at least one negative eigenvalue, then there exit causal (temporal) correlations and PDM represents the time-like correlated system state. In other words, the correlations are purely temporal as they are obtained for a system between time $t_1$ and the same system at a later time $t_2$ evolved through the channel $\mathcal{E}_{B \leftarrow A}$. When PDM is positive semi-definite, it is equivalent to the state of a space-like correlated systems. Hence, PDM formalism accounts for spatial and temporal correlations in unified way. 
Based on the above considerations, a two-point PDM under any channel $\mathcal{E}$ can be given an alternative representation \cite{pisarczyk2019causal}:
\begin{align}
\mathcal{P_{\rm AB}} = \left(I \otimes \mathcal{E} \right)\left[\bigg\{\rho \otimes \frac{I}{2}, Q_{\mathsf{swap}} \bigg\}\right]
\label{eq:pdm2}
\end{align}
where $Q_{\mathsf{swap}} := \frac{1}{2} \sum_{i=0}^{3} \sigma_i \otimes \sigma_i$ and $\{\hat{a},\hat{b}\} = \hat{a}\hat{b} + \hat{b}\hat{a}$ is the anti-commutator of operators $\hat{a}$ and $\hat{b}$, and $\sigma_i$ are Pauli-X,Y and Z operators with $\sigma_0=I$. Throughout the paper we consider one time use of channel acting on the initial qubit state $\rho$ with measurements before and after the use of the channel.  This essentially creates two-point temporal correlations \cite{zhao2018geometry}. E.g., for an identity channel $\mathcal{E}= I$ with an initial state set to $\rho=\frac{I}{2}$, the PDM is obtained as:
\begin{align}
\mathcal{P}^{\rm \small AB}= \left(
\begin{array}{cccc}
\frac{1}{2} & 0 & 0 & 0 \\
0 & 0 & \frac{1}{2} & 0 \\
0 & \frac{1}{2} & 0 & 0 \\
0 & 0 & 0 & \frac{1}{2} \\
\end{array}
\right)
\end{align}
Note that it has eigenvalues $\left\{-\frac{1}{2},\frac{1}{2},\frac{1}{2},\frac{1}{2}\right\}$ which implies that a PDM generally may not be positive semi-definite.  In this work we consider a simple case of two-point correlations obtained out of 2 sequential measurements before and after the single use of the channel. Interestingly, through Choi-Jamiolkowski (CJ) isomorphism, Eq. (\ref{eq:pdm2}) in fact can be represented as 
\begin{align}
\mathcal{P}_{\rm AB} = \{\rho_A \otimes \frac{I}{2}, \chi_{\rm AB}\},
\label{eq:jordan}
\end{align}
where 
\begin{align}
\chi_{\rm AB} := \sum_{i,j} (I_{A} \otimes \mathcal{E}_{B \leftarrow A}) (\ket{i}\bra{j}_A \otimes \ket{j}\bra{i}_B),
\label{eq:choi}
\end{align} is a state that is CJ isomorphic to the channel $\mathcal{E}_{B \leftarrow A}$. The above Eq. (\ref{eq:jordan}) is known as the Jordan product representation. This feature has been used to relate the causality measure with the quantum capacity of the channel \cite{pisarczyk2019causal}.

\section{Causality monotone and divisibility of dynamical map \label{sec:3}}
It is known that under any local operations and classical communication (LOCC) protocol, (spatial) entanglement shared between a two parties can not increase - one of the properties on which the definition of an entanglement monotone relies. Local operations on the other hand are generally completely positive trace preserving (CPTP) or trace-non-increasing maps representing devices that effect a transformation on the system state. They are given an operators-sum representation: given a local operation $\mathcal{E}$, we have $\mathcal{E}[\mathcal{\rho}] = \sum_j K_j \mathcal{\rho} K^\dagger _j$, where are $K_j$ are the Kraus operators of the channel $\mathcal{E}$ acting on the state $\mathcal{\rho}$. PDM can be used to define a causality monotone which decreases under a CPTP Markovian channel. We will see shortly that this assumption is violated for a non-Markovian (or CP-indivisible) map. However, note that classical communication must be excluded since it induces causal relationships, therefore not accounted for in order to define the causality monotone.

A causality measure, based on PDM $\mathcal{P}$ was defined in \cite{fitzsimons2015quantum} as
\begin{align}
f_{\rm cm} =  \| \mathcal{P}\|_1 -1,
\label{eq:fcm}
\end{align} where $\|\mathcal{P}\|_1$ is the trace norm of the PDM given by $ \| \mathcal{P}\|_1 = \text{Tr}[\sqrt{\mathcal{P}^\dagger \mathcal{P}}]$. Here, $\mathcal{P}^\dagger$ is the Hermitian conjugate of $\mathcal{P}$. The function $f_{\rm cm} $ satisfies all the necessary conditions for it to be a monotone. Most notably it is monotonically decreasing function for CPTP maps. Note that using Stinespring dilation theorem, a local quantum operation $\mathcal{E}$ can be represented as a unitary operation on an extended Hilbert space and since trace norm is known to monotonically decrease under partial trace, any CPTP map is contractive under trace norm \cite{peres-garcia2006contractivity}. Here we make use of more generalized causality measure introduced in \cite{pisarczyk2019causal}, given by 
\begin{align}
F:= \log_2(f_{\rm cm} + 1) = \log_2\|\mathcal{P}_{\rm AB}\|_1,
\label{eq:measure}
\end{align}
It is shown  that for a maximally mixed input state $\rho = \frac{I}{2}$ in Eq. (\ref{eq:pdm2}), the function $F$ reduces to \cite{zhao2018geometry}  \begin{align}
\mathcal{N} :=  \log_2\|(\chi_{\rm \small AB})^{\rm PT}\| _1,
\label{eq:negativity}
\end{align} 
where $PT$ denotes positive-partial-transpose. Here, $\chi_{\rm AB}$ is the Choi matrix given in Eq.(\ref{eq:choi}). \bla Put in words, Eq. (\ref{eq:negativity}) is the logarithmic negativity \cite{vidal2002computable} of the Choi state of the channel $\mathcal{E}$.

Certain properties of $F$ are worth mentioning. $F=0$ if $\mathcal{P}$ is positive semidefinite and $F=1$ for $\mathcal{P}_2$ obtained from two consecutive measurements on a single
qubit closed system, which is when the initial state $\rho$ is not interacting with the external environment, which is equivalent to saying that  $\mathcal{E}=U$, where $U$ is some unitary. 

The map $\mathcal{E}$ is said to be CP-divisible in a given interval $\{t+\tau, t_0\}$ when the condition
\begin{align}
\mathcal{E}(t+\tau,t)\mathcal{E}(t,t_0) = \mathcal{E}(t+\tau,t_0)
\label{eq:cp-div}
\end{align}
is satisfied \cite{RHP10}, where $t_0 \le t \le t+\tau$. As said earlier, $F$ is non-increasing for local quantum operations, for a Markovian channel $\mathcal{E}$, the following condition holds:
\begin{align}
F_{\mathcal{E}[\rho]}(t) \ge F_{\mathcal{E}[\rho]}(t+\tau),
\label{eq:monotonicity}
\end{align}
with $t+\tau \ge t$. As noted in Section (\ref{sec:pdm}), PDM formalism has natural operational interpretation in the sense that it requires two experimental runs to determine the causality measure; i.e., the qubit is measured before and after the evolution down a channel. Consider that the qubit evolves from time $t_0$ to $t$ via the channel $\mathcal{E}(t,t_0)$, then PDM is obtained by making measurement at initial time $t_0$ and final time $t$ and the corresponding causality measure at $t$ is obtained as in Eq.(\ref{eq:measure}). Now the same qubit evolves under the channel $\mathcal{E}(t+\tau,t)$ and the PDM is obtained by just making a measurement at time $t+\tau$, because it has been measured already at time $t$, which is evident from the continuity of evolution in Eq.(\ref{eq:cp-div}) giving rise to continuity of $F$ in Eq.(\ref{eq:monotonicity}). Hence, a single time argument is used for representation. Therefore, determination of the condition of monotonicity in Eq. (\ref{eq:monotonicity}) essentially requires three experimental runs.

It is in fact straightforward to generlaize the PDM formalism to $k$-time instances. Interestingly, it was shown in \cite{zhang2020quantum} that a $3$-time PDM can be suitably mapped to $2$-switch which gives rise to processes with indefinite causal order. In \cite{milz2018entanglement}, it was noted that every causally unordered process, such as those studied using process matrices \cite{oreshkov2012quantum}, can be realized by a non-Markovian dynamics as described in the process tensor framework \cite{pollock2018non,pollock2018operational}. In \cite{zhang2020quantum} it was also shown that a process matrix can be mapped to PDM in four different ways. Here, our work is motivated to realize a measure, similar in spirit to \cite{breuer2009measure}, that quantifies dynamical memory in a given physical process and a two-time PDM is used for simplicity.

It is known that non-Markovian dynamics may give rise to information back-flow incurring revivals of quantum resources back to the system from the environment.  Since $F$ is a monotone, violation of condition (\ref{eq:monotonicity}) hence of (\ref{eq:cp-div}) implies that monotonicity will be broken under a CP-indivisible channel hence signal non-Markovianity through revivals of temporal quantum correlations.

\section{Non-Markovianity measure based on causality monotone}

Based on the above considerations, one can define a measure of non-Markovianity using causality monotone $F$ for channel $\mathcal{E}$ acting on an initial state $\rho$ as below:
\begin{align}
\mathcal{M} := \underset{\rho}{\rm max} \int_{\sigma_{(\rho,\mathcal{E},t)}>0}  dt \; \sigma_{(\rho,\mathcal{E},t)}
\label{eq:cmeasure0}
\end{align}
where $$\sigma_{(\rho,\mathcal{E},t)}= \frac{dF}{dt}.$$ Here the integration is done over positive slope of $F$. Following \cite{RHP10,chen2016quantifying}, one may make use of an equivalent definition given by
\begin{align}
\mathcal{M} := \underset{\rho}{\rm max} \int_{t_0}^{t_{\rm max}} \biggl| \frac{dF}{dt} \biggr|  dt + (F_{t_{\rm max}} -  F_{t_0} ).
\label{eq:cmeasure}
\end{align}
The integral (\ref{eq:cmeasure}) is such that for a non-Markovian process, the derivative of $F$ is positive and $\mathcal{M} > 0$. And for a time-dependent Markovian process, the derivative of $F$ is negative and hence $\mathcal{M}=0$. 
The channel becomes CP-indivisible when $\sigma_{(\rho,\mathcal{E},t)}>0$, which we will show later that it corresponds to negativity of decay rate in the canonical master equation representing the underlying dynamics.

It is shown \cite{fitzsimons2015quantum} that $f_{\rm cm}$ satisfies the convexity condition $f_{\rm cm}(\sum_{j} p_j \mathcal{P}_j) \le  \sum_{j} p_j f_{\rm cm}(\mathcal{P}_j)$. Since it is monotonic under  CPTP maps, one may define a measure of non-Markovianity based on $f_{\rm cm}$ also. But $f_{\rm cm}$ is not additive in which case the relation with the quantum capacity may not be possible for the non-Markovianity measure based on $f_{\rm cm}$. Whereas $F$ is additive and therefore, convexity of $F$ can be sacrificed in favor of additivity which will in fact be beneficial to relate the resulting causality measure (\ref{eq:measure}) with quantum capacity through Choi-Jamiolkoswski isomorphism \cite{pisarczyk2019causal}. Note the similarity of $F$ with logarithmic negativity of a bipartite entangled state, which is a well known entanglement monotone used to define a measure of non-Markovianity \cite{RHP10}. Interestingly, for a maximally mixed initial state (i.e., $\rho=\frac{I}{2}$), Eq. (\ref{eq:measure}) gives exactly the same results as does Eq. (\ref{eq:negativity}) - the logarithmic negativity of the Choi matrix of the map $\mathcal{E}$ - in quantifying non-Markovianity of quantum dynamics, and it has been shown \cite{pisarczyk2019causal} to upper bound the quantum capacity of the channel $\mathcal{E}$. 

The method may be extended to $n$ parallel use of non-Markovian channel $\mathcal{E}$ so that one may relate the causality measure to quantum capacity thereby allowing for a more general capacity based non-Markovianity measure. The measure does not require optimization in such case (i.e., with a maximally mixed initial state), else requires optimization over all initial input states. We may omit the discussion in this work. But, it is pertinent to point out that for the causality measure to be an upper bound on quantum capacity, it was assumed \cite{pisarczyk2019causal} that the sequential use of channels is uncorrelated. Because, correlated quantum channels bring in classical memory effects \cite{kretschmann2005quantum} and hence one can \textit{not} consider $n$ parallel use of channels as a tensor product $\mathcal{E}^{\otimes n}$. Extending the work of \cite{pisarczyk2019causal} to memory channels \cite{caruso2014quantum} is an interesting open problem.

\section{Example 1: The damped Jaynes-Cummings model}
Let us consider damped Jaynes-Cummings (JC) model \cite{breuer2016colloquium} of a two level system interacting with a dissipative bosonic reservoir at zero temperature. The total system-reservoir Hamiltonian is
\begin{align}
H_{\rm \small tot} &= \frac{\omega_0 \sigma_z}{2} + \sum_j \omega_j a_j^\dagger a_j + \sum_j (g_j \sigma_{+} a_j + g^\ast _j \sigma_{-} a^\dagger _j)
\end{align}
where $a_j^\dagger$ and $a_j$ are creation and annihilation operators, and $g_j$ are coupling parameters and $\sigma_{+}=\vert 1 \rangle \langle 0 \vert$ and $\sigma_{-}=\vert 0 \rangle \langle 1 \vert$. The above model has a Lorentian spectral density $J(\omega) = \gamma_0 b^2/2 \pi [(\omega_0 - \Delta - \omega)^2 + b^2]$,
where $ \Delta = \omega-\omega_0 $ is the detuning parameter which governs the shift in frequency $\omega$ of the qubit from the central frequency $ \omega_0$ of the bath. $ \omega_0$ represents the energy gap between ground state $\ket{0}$ and exited state $\ket{1}$. $\gamma_0$ quantifies the strength of the system-environment coupling and $b$ the spectral bandwidth. The reduced dynamics of the qubit interacting with such an environment can be modeled as an amplitude damping (AD) channel: $\mathcal{E}^{\rm \small AD}[\rho] = \sum_i A_i \rho A_i^\dagger$ with the Kraus operators
\begin{align}
A_1 = \left(
\begin{array}{cc}
1 & 0 \\
0 & \sqrt{1-r(t)} \\
\end{array}
\right) \quad;\quad  A_2 = \left(
\begin{array}{cc}
0 & \sqrt{r(t)} \\
0 & 0 \\
\end{array}
\right).
\label{eq:adkraus}
\end{align}
 In this work, we will consider the case of no detuning ($\Delta=0$) i.e., when the system qubit is in resonance with the central frequency modes of the environment  i.e., $\omega_0=\omega$. \bla Here then, for damped JC model, $r(t)$ takes the form $r(t)=1-|G(t)|^2 $, with
 \begin{align}
G(t) =e^{-\frac{b t}{2}} \left( \frac{b}{d} \sinh \left[\frac{d t}{2}\right] + \cosh \left[\frac{d t}{2}\right] \right),
\label{eq:G}
\end{align}
where $d = \sqrt{b^2 - 2\gamma_0 b}$. \bla The map (\ref{eq:adkraus}) satisfies the completeness condition $\sum_{j=0}A_j^\dagger A_j = I$. 

Before going further, note that the operational meaning of the CP-divisibility condition (\ref{eq:cp-div}) and the monotonicity condition (\ref{eq:monotonicity}) were presented in Section (\ref{sec:3}) with 3 time instances and 3 measurements on the qubit. In the following calculations, we make one use of quantum channel which takes the qubit from time $t_0$ to $t$, which in fact constitutes a continous "family" of channels, which suffices for our purpose here. This is done for the simplicity of the presentation of the measure, however calculating it for higher dimensional PDM with multiple time measurements would be more involved but straightforward.

\begin{figure}
	\centering
	\includegraphics[width=0.4\textwidth]{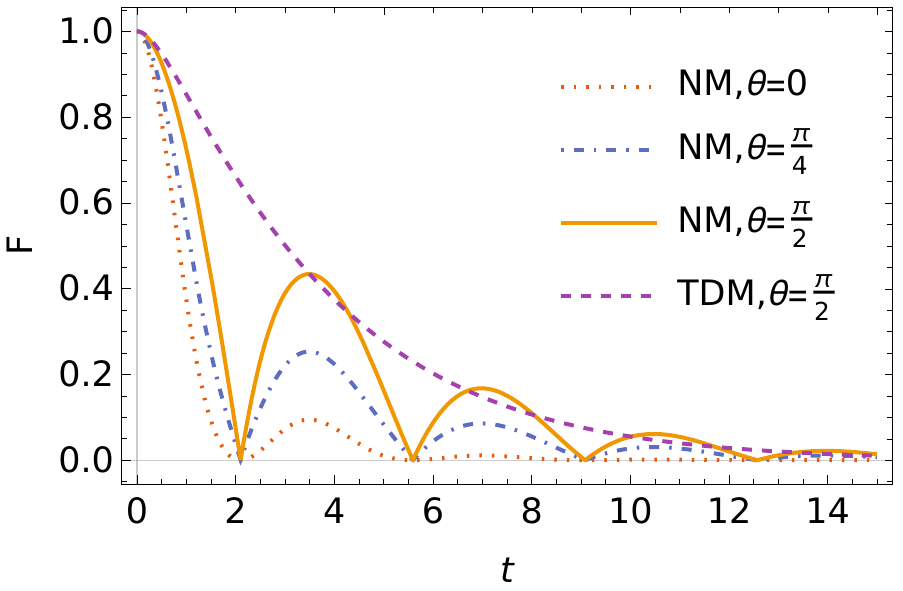}
	\caption{(Color online.) Breakdown of monotonicity of causality measure (\ref{eq:measure}) for AD for a given range of time. For the values $\gamma_0=3$ and $b=0.6$, the curves with $\theta=0$ (dotted, yellow), $\theta=\frac{\pi}{4}$ (dot-dashed, blue), and $\theta=\frac{\pi}{2}$ (bold, orange) represent non-Markovian (NM) processes with revivals of temporal (causal) correlations. And for $\gamma_0=0.6$ and $b=3$ the dashed purple curve, for an  initial state with $\theta=\frac{\pi}{2}$, represents time-dependent Markovian (TDM) process, where the correlations fall monotonically.} 
	\label{fig:pdmad}
\end{figure}

\begin{figure}[ht!]
	\centering
	\includegraphics[width=0.4\textwidth]{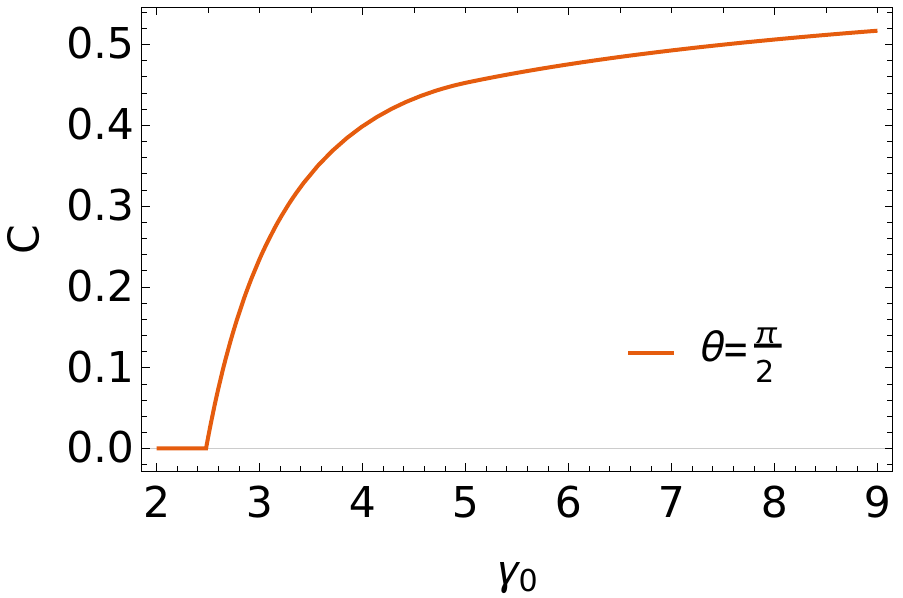}
	\caption{(Color online.) Normalized measure $C$ of non-Markovianity (\ref{eq:cmeasure}) for AD, in its non-Markovian regime given by $\frac{\gamma_0}{b} > \frac{1}{2}$, for $\Delta=0$, against the coupling strength $\gamma_0$, with a fixed $b=1.1$. The integration interval in Eq. (\ref{eq:cmeasure}) is chosen to be $t=0$ to $t_{\rm max}=2$, and the initial state was set to $\rho=\vert 0 \rangle \langle 0 \vert$, i.e., $\theta=\frac{\pi}{2}$ for which the measure attains a maximum, which can be verified from Fig.(\ref{fig:pdmad}) where for $\theta=\frac{\pi}{2}$ PDM attains a maximum recurrence and that the measure the above Figure is defined over the positive slope of the bold, orange curve in Fig.(\ref{fig:pdmad}) with $\theta=\frac{\pi}{2}$.} 
	\label{fig:cmeasuread}
\end{figure}

Now, consider a general state 
\begin{equation}
\rho = \left(
\begin{array}{cc}
\sin ^2(\theta ) & \frac{1}{2} \sin (2 \theta ) \\
\frac{1}{2} \sin (2 \theta ) & \cos ^2(\theta ) \\
\end{array}
\right).
\end{equation}From Eq. (\ref{eq:pdm2}) for an initial pseudo-pure state $\rho=\vert 0 \rangle \langle 0 \vert$ which is obtained when $\theta=\frac{\pi}{2}$, one obtains PDM for the above channel as below:
\begin{align}
\mathcal{P}^{\rm \small AD} = \left(
\begin{array}{cccc}
1 & 0 & 0 & 0 \\
0 & 0 & \frac{\sqrt{1-r(t)}}{2} & 0 \\
0 & \frac{\sqrt{1-r(t)}}{2} & 0 & 0 \\
0 & 0 & 0 & 0 \\
\end{array}
\right)
\label{eq:pdmad}
\end{align}
which has the eigenvalues $e_1=-\frac{\sqrt{1-r(t)}}{2}, \, e_2=\frac{\sqrt{1-r(t)}}{2},\,e_3=1,\,e_4=0$. Note that $e_1$ is always negative indicating that $\mathcal{P}^{\rm \small AD}$ is not positive semi-definite. The measure (\ref{eq:measure}) now reads
\begin{align}
F^{\small \rm AD} =\log_2 \left(1+\sqrt{1-r(t)}\right) 
\label{eq:cmeasuread}
\end{align} which is plotted in Fig. (\ref{fig:pdmad}) for the considered noise in its Markov and non-Markov regimes.  Note that, for the particular case of initial state $\rho=\ket{0}\bra{0}$, when $G(t)=\sqrt{1-r(t)}= 1$, then $F=1$. That is when there is no noise acting on the system, the temporal correlations are maximum. And when $r(t)=1$, i.e., when the system is maximally damped, both $e_1=e_2=0$, hence the causality measure $F=0$, which implies that causal correlations die out at that point.  One may observe the revivals of causal correlations which happens when $G(t)$ becomes negative; see Fig. (\ref{fig:pdmad}). When all the eigenvalues are positive, then the PDM is called \textit{acausal} in the sense that there are no causal relationships, and if at least one eigenvalue is negative then $F>0$ hence there exist causal correlations and the PDM then is called \textit{causal}. Note that when the system interacts with the environment in weak-coupling regime, the causal correlations die slower than that compared with in the strong-coupling regime. The reason is that the stronger the coupling, the rapid the process. Interestingly, strong-coupling regimes make PDM to become acausal at times, but because of non-Markovian nature, the correlations can revive making PDM causal again. The normalized measure $C$, normalized as $C = \frac{\mathcal{M}}{1+\mathcal{M}}$ from Eq. (\ref{eq:cmeasure}) for AD, such that $0 \le C \le 1$, is plotted in Fig. (\ref{fig:cmeasuread}). 

\section{Comparison with some of the existing measures of non-Markovianity}

\subsection{Comparison with the decay rate measure} A measure of non-Markovianity based on canonical decay rate was given by Hall-Cresser-Li-Anderson (HCLA) in \cite{hall2014canonical}. The canonical master equation $\dot{\rho}(t) = \mathcal{L}[\rho(t)]$ of the process considered in (\ref{eq:adkraus}) is given by \begin{align}
\frac{d \rho_s(t)}{dt} =
\gamma(t) [\sigma_- \rho_s(t) \sigma_+ - \frac{1}{2}\{\sigma_+ \sigma_-, \rho_s(t)\}], 
\label{eq:mastereq}
\end{align} 
where 
$\gamma(t) = - 2 \Re[\frac{\dot{G}(t)}{G(t)}]$ is the time-dependent decay rate, and $G(t)$ is the decoherence function given Eq.(\ref{eq:G}). Given that, the HCLA measure is given as $\mathcal{N}_{\rm \small{HCLA}} =  \int_{0}^{t_{\rm max}}- \gamma(t) dt$, where the integral is over only the negative parts of $\gamma(t)$. Now, as was noted earlier, the measure is defined over positive slope of $\frac{dF}{dt}$, which, from Eq. (\ref{eq:cmeasuread}), is found to be 
\begin{align}
\frac{dF}{dt} = \gamma(t).\frac{ G(t)}{2(1+\sqrt{G^2(t)})}
\label{eq:comp}
\end{align}
where $\gamma(t)= 2 \Re\bigg(\frac{\gamma_0 }{\sqrt{1-\frac{2\gamma_0}{b}} \coth \left(\frac{1}{2} bt \sqrt{1-\frac{2 \gamma_0}{b}}\right)+1}\bigg)$ is the decay rate with $\Delta=0$. Note that the denominator of (\ref{eq:comp}) is always positive. The fraction $\frac{dF}{dt} > 0$ when $G(t) < 0$ i.e., when the revivals occur, and that is when $\gamma(t) < 0$. Therefore, negativity of decay rate gives rise to CP-indivisibility of $\mathcal{E}$ which is when the monotonicity condition (\ref{eq:monotonicity}) of causality measure $F$ will be broken. The measure of non-Markovianity according to negativity of decay rate is simply $\mathcal{N}_{\rm HCLA} = -\int_{0}^{t_{\rm max}} \gamma(t) dt$. 

Observe that the time-local generator $\mathcal{L}$ has infinite singularities. At every instance when the generator becomes singular, the dynamical map $\mathcal{E}^{\rm \small AD} = T \exp \{- \int_{t_0}^{t}\mathcal{L}(s) ds\}$ becomes momentarily non-invertible. Here, $t \ge s \ge t_0$ and $T$ is the time-ordering operator. Interestingly, at each such point in time the PDM becomes positive semi-definite, i.e., particularly for the considered example with $\rho =\ket{0}\bra{0}$, the eigenvalues become: $e_1(t)=e_2(t)=e_4(t)=0$ and $e_3(t) = 1$. In other words, the causality measure $F=0$ at each such points (see Fig. (\ref{fig:pdmad})). One observes revivals of causal correlations after each singularity when the decay rate becomes negative. It would an interesting future direction to look at the relationship between singularities in a dynamical map \cite{hou2012singularity}  and correlations in space-time. 

\subsection{Comparison with trace distance.} One may consider comparing the measure (\ref{eq:measure} and \ref{eq:cmeasure0}) with trace distance between a pair of initial states $\{\rho_1,\rho_2\}$ under a channel $\mathcal{E}$, given by $\mathcal{D}(\rho_1,\rho_2)=\frac{1}{2}\text{tr}| \mathcal{E}[\rho_{1}]-\mathcal{E}[\rho_{2}]|$. The distinguishability of a pair of initial states undergoing a CPTP map decreases over time. It can be shown \cite{breuer2009measure} that $\ket{+}$ and $\ket{-}$ are an optimal pair for an AD process given in Eq. (\ref{eq:adkraus}). However, for a qubit undergoing AD, if the temporal correlations are set up, meaning that the future qubit under AD gets correlated with itself in the past such that both qubit states (in time) can not be written in a tensor product form, then it is seen in this work that the initial state $\ket{0}$ maximizes the measure (\ref{eq:cmeasure}) for AD (See, caption in Fig. (\ref{fig:pdmad})). Interestingly, TD under AD for initial pair of states $\ket{+}$ and $\ket{-}$ reads $\mathcal{D}(\rho_{+},\rho_{-})=\sqrt{1-r(t)}$, which is exactly equal to the causality measure (\ref{eq:fcm}) for an input state $\ket{0}$ under AD i.e., $f_{cm} = \sqrt{1-r(t)}$. Hence we have
\begin{align}
F_{\rho = \vert 0 \rangle \langle 0 \vert} = \log_2 (1 + \mathcal{D}(\rho_{+},\rho_{-})),
\end{align}where $\rho_{\pm} =\vert \pm \rangle \langle \pm \vert$ are the initial pair of orthogonal states and $\rho =\vert 0 \rangle \langle 0 \vert$ is the initial input of the PDM. For different input states to PDM, the relationship might be more involved.


One may say that revival of temporal correlations can be interpreted as equivalent to information back-flow via recurrence of trace-distance, given that the measures under comparison are optimized over initial input states. Therefore, one must keep in mind the optimization over input states while comparing the causality based non-Markovianity measure with that of others.

\section{Example 2: non-Markovian generalized amplitude damping channel \label{sec:example2}}
As a second example, let us consider non-Markovian generalized amplitude damping channel (GADC) introduced in \cite{liu2013nonunital}. This example is given in order to exemplify a feature of the causality measure that it can detect non-Markovianity of the channel while trace distance doesn't when memory effects are coming solely from non-unital part. 

In \cite{liu2013nonunital}, it was pointed out that trace distance may fail to detect non-Markovianity for non-unital channels. Here, we show that causality measure detects memory originating from both unital and non-unital parts.  We consider  a generalized amplitude damping (GAD) channel given by the Kraus operators representation \cite{srikanth2008squeezed,liu2013nonunital}
\begin{align}
E_{1}&=\sqrt{1-p(t)}\left[\begin{array}{cc}
1 & 0\\
0 & \sqrt{1-\lambda(t)}
\end{array}\right]; \nonumber \\  E_{2}&=\sqrt{1-p(t)} \left[\begin{array}{cc}
0 & \sqrt{\lambda(t)}\\
0 & 0
\end{array}\right]; \nonumber \\
E_{3}&=\sqrt{p(t)}\left[\begin{array}{cc}
\sqrt{1-\lambda(t)} & 0\\
0 & 1
\end{array}\right]; \nonumber \\ E_{4}&=\sqrt{p(t)}\left[\begin{array}{cc}
0 & 0\\
\sqrt{\lambda(t)} & 0
\end{array}\right].
\label{eq:gadkraus}
\end{align}
with $p(t) = \sin^2(\omega t)$ and $\lambda(t) = 1- e^{-t}$.

From Eq. (\ref{eq:pdm2}), for an initial state $\rho = \ket{0}\bra{0}$, one obtains PDM for GAD as below
\begin{align}
\mathcal{P}^{\small \rm GAD} = \left(
\begin{array}{cccc}
1-p \lambda  & 0 & 0 & 0 \\
0 & p \lambda  & \frac{\sqrt{1-\lambda }}{2} & 0 \\
0 & \frac{\sqrt{1-\lambda }}{2} & 0 & 0 \\
0 & 0 & 0 & 0 \\
\end{array}
\right)
\label{eq:pdmgad}
\end{align}
which has the eigenvalues $h_1=0,\,h_2=(1-\lambda  p),$ and $h_{3,4}=\frac{1}{2} \left( \lambda  p \pm \sqrt{1 -\lambda +\lambda ^2 p^2}\right)$. Now the measure (\ref{eq:measure}) for GAD reads 
\begin{align}
F^{\small \rm GAD} = \log_2 \left(\sqrt{1+2 x+y}+\sqrt{1-2 x+y}+2 z\right)-1 . 
\end{align}
where $x:=\sqrt{\lambda ^2 p^2 \left(\lambda  \left(\lambda  p^2-1\right)+1\right)}$ and $y:=\lambda  \left(2 \lambda  p^2-1\right)$ and $z :=\sqrt{(\lambda  p-1)^2}$. In the above expressions we have used the notation $\lambda(t) = \lambda$ and $p(t)=p$, for simplicity. 

\begin{figure}
	\centering
	\includegraphics[width=0.4\textwidth]{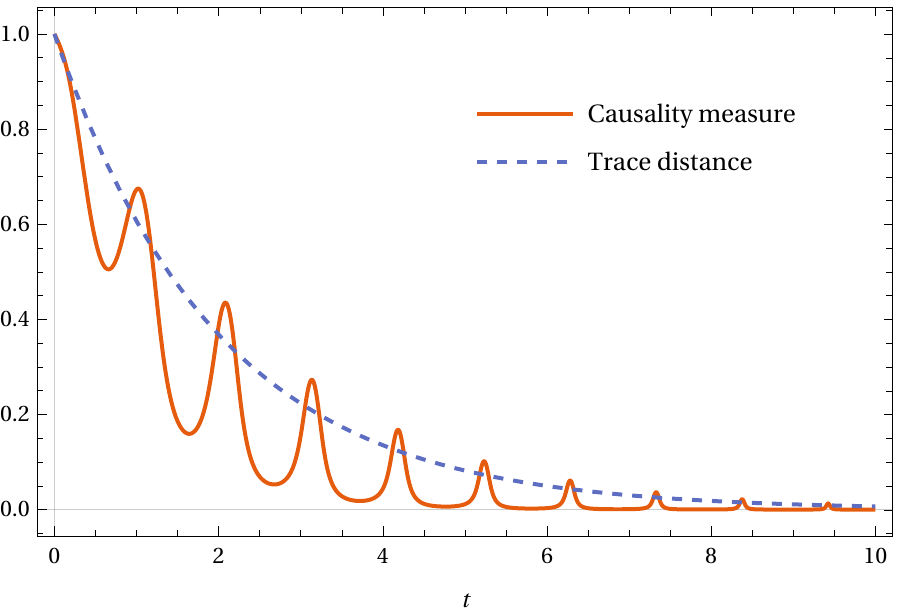}
	\caption{(Color online.) Breakdown of monotonicity of causality measure (\ref{eq:measure}) for $\omega=3$, (bold, red curve) and the behavior of trace distance (TD) under non-Markovian GAD \cite{liu2013nonunital}.}
	\label{fig:pdmgad}
\end{figure}

Note the revivals of temporal correlations for $\omega=3$ in Figure (\ref{fig:pdmgad}), which is evident from the observation that $p(t)$ appears in the measure. Whereas, the trace distance (TD) between a pair of initial states $\ket{+}$ and $\ket{-}$ for the channel (\ref{eq:gadkraus}) evaluates to $\sqrt{1-\lambda(t)}$ (plotted in Fig. (\ref{fig:pdmgad})) which is independent of $p(t)$. It can be shown that TD is independent of $p(t)$ for all initial pairs of orthogonal states under the action of GAD. 

Further, note that quantum (spatial) correlation, such as entanglement, based measures such as given in \cite{RHP10,luo2012quantifying,liu2013nonunital,fanchini2014non-Markovianity,haseli2014non-Markovianity}, do detect non-Markovianity of the above channel (\ref{eq:gadkraus}). Although, it is important to note the difference between these measures and the current one. Our measure is based on causal correlations between the \textit{same} system measured at two different times $t_1$ and $t_2$ where the state of the system at $t_2$ is fixed by the evolution effected by the channel $\mathcal{E}$. It is known that RHP measure is equivalent to decay rate (or HCLA) measure by a factor of $\frac{2}{d}$, where $d$ is the dimension of the system, i.e., $\mathcal{N}_{\rm RHP} = \frac{2}{d}\mathcal{N}_{\rm HCLA}$ \cite{hall2014canonical,shrikant2018non-markovian}. Since the comparison of the measure introduced in this paper has been done with the decay rate measure, the comparison between entanglement based measures and the causality based measure is straightforward, hence we do not propose to do it here. 

One may obtain a measure of non-Markovianity using Eq. (\ref{eq:cmeasure}), which is not done here, since the purpose of this example was to show that the causality measure can detect non-Markovianity of non-unital channels.

\section{Conclusions}
A measure of non-Markovianity is introduced based on causality monotone given in Eq. (\ref{eq:measure}) arising out of pseudo-density-matrix (PDM) formalism \cite{fitzsimons2015quantum}. A more generalized monotone \cite{pisarczyk2019causal}, which is just logarithm of the trace norm of PDM, is used to define the measure. The function $F$ in Eq. (\ref{eq:negativity}) is non-increasing for local operations which are CPTP maps in general. Breakdown of monotonicity of $F$ is an indication that the corresponding map is CP-indivisible. A suitable quantifier of non-Markovianity is given in Eq. (\ref{eq:cmeasure}). 


Its comparison with the decay rate measure for the case of damped JC model is also done showing that negativity of decay rate entails breakdown of monotonicity of causality measure hence of CP-divisibility of a channel. Since causality measure, for certain input states of PDM, is shown to upper bound the quantum capacity of a channel, it can be said that it upper bounds the capacity based measures introduced \cite{bylicka2014non}. 


Although the PDM formalism, as proposed, is for qubits, it is important to note that it \textit{can be} generalized to the systems of arbitrary dimensions across time \cite{fitzsimons2015quantum,horsman2017can}. Therefore, the measure presented in this work should also apply to systems of arbitrary finite dimension, though it is exemplified with a system of dimension 2 in the present paper.  The framework, as it stands, is also for multiple qubits in space-time, hence use of, say $n$ parallel use of qubit channels is possible, which in fact enables one to quantify quantum capacity of a channel using causality measure \cite{pisarczyk2019causal}. 

Interesting future direction would be to investigate its applicability to various models of open system dynamics and also compare it with the recently introduced non-Markovianity measure based on temporal steering \cite{chen2016quantifying}, and with that of other spatial correlation based measures such as entanglement \cite{RHP10}, accessible information \cite{fanchini2014non-Markovianity}, mutual information \cite{luo2012quantifying}. It is important to note that causal correlations due to PDM contain the strongest form of quantum direct cause and it was shown \cite{ku2018hierarchy} that temporal steering can be a weaker form of quantum direct cause. In this light, non-Markovianity measure proposed in this work can be conjectured to be stronger than that proposed using temporal steerable weight \cite{chen2016quantifying}. However, the proof is left for a future work.

It is worth noting that treating quantum mechanics beyond a definite causal structure is under intense investigation \cite{zhang2020quantum}. As is known, PDM contains correlations without definite causal order, it is evident from this work that revival of causal correlations due to (non-Markovianity) CP-indivisibility brings back indefiniteness of causal order. 

\section*{Acknowledgments}
The author thanks R. Srikanth for insightful discussions and suggestions and also S. Aravinda and Ranjith V. for helpful comments on the draft. This work was partially supported by the Department of Science and Technology (DST), Govt. of India,  through  the  project EMR/2016/004019.

\end{document}